\documentclass[twocolumn,prb,showpacs,superscriptaddress,amsmath,citeautoscript,floatfix]{revtex4}
\usepackage{graphicx}
\usepackage{bm}
\usepackage{amssymb}

\begin{document}

\title{Temperature dependent  nucleation and annihilation of individual magnetic vortices}

\author{G.\ Mihajlovi\'{c}}
\email{mihajlovic@anl.gov}
\affiliation{Materials Science Division, Argonne National Laboratory, Argonne, IL 60439\\}
\author{M.\ S.\ Patrick}
\affiliation{Materials Science Division, Argonne National Laboratory, Argonne, IL 60439\\}
\author{J.\ E.\ Pearson}
\affiliation{Materials Science Division, Argonne National Laboratory, Argonne, IL 60439\\}
\author{V.\ Novosad}
\affiliation{Materials Science Division, Argonne National Laboratory, Argonne, IL 60439\\}
\author{S.\ D.\ Bader}
\affiliation{Materials Science Division, Argonne National Laboratory, Argonne, IL 60439\\}
\affiliation{Center for Nanoscale Materials, Argonne National Laboratory, Argonne, IL 60439\\}
\author{M.\ Field}
\affiliation{Teledyne Scientific Company LLC, Thousand Oaks, CA 90360\\}
\author{G.\ J.\ Sullivan}
\affiliation{Teledyne Scientific Company LLC, Thousand Oaks, CA 90360\\}
\author{A.\ Hoffmann}
\affiliation{Materials Science Division, Argonne National Laboratory, Argonne, IL 60439\\}
\affiliation{Center for Nanoscale Materials, Argonne National Laboratory, Argonne, IL 60439\\}

\date{\today}

\pacs{75.60.Jk, 75.75.-c, 07.55.Jg}

\begin{abstract}
We studied the temperature dependence of the magnetization reversal in individual submicron permalloy disks with micro-Hall and bend-resistance magnetometry. The nucleation field exhibits a nonmonotonic dependence with positive and negative slopes at low and high temperatures, respectively, while the annihilation field monotonically decreases with increasing temperature, but with distinctly different slopes at low and high temperatures. Our analysis suggests that at low temperatures vortex nucleation and annihilation proceeds via thermal activation over an energy barrier, while at high temperatures they are governed by a temperature dependence of the saturation magnetization.
\end{abstract}

\maketitle

Significant experimental and theoretical efforts have been devoted over the last decade to understand magnetization reversal and dynamics of vortices in submicron ferromagnetic disks
\cite{Guslienko2008, Antos2008}.  The relatively simple geometry makes magnetic vortices an ideal model system to understand how different interactions ({\em e.g.}, exchange {\em vs.}\ dipolar) contribute to their magnetic behavior.  Many of the features of magnetic vortices can be understood with analytical models.  For example, the energy barriers for the zero-field transitions between single-domain and vortex configurations have been calculated and successfully compared to experimentally observed thermally activated transitions \cite{Ding2005}.  However, the problem of energy barriers between vortex and uniform single-domain states and the related problem of the temperature $T$ dependence of the critical fields for vortex nucleation $H_n$ and annihilation $H_{an}$ remain relatively unexplored \cite{Guslienko2008}. Theoretically $H_n$ and $H_{an}$ are derived within the rigid-vortex model that does not include thermal effects \cite{Guslienko2001}. However, $T$ dependencies of $H_n$ and $H_{an}$ have been observed in submicron square permalloy (Py) structures \cite{Li2001}, in which the magnetization reversal also proceeds via nucleation and annihilation of vortices, as well as in sub-100 nm Fe disks \cite{Dumas2007}. In addition, in micron-size Py disks, the remanence, the coercive field, and the initial susceptibility are all found to be $T$ dependent \cite{Shima2002}. These studies were performed by measurements of large arrays of disks. Thus, they reflect a statistical average of the reversal modes of individual disks, which, even when the disks are nominally identical, can significantly differ from each other \cite{Rahm2003}. This prevents more detailed insights into the physics of the magnetization reversal and may obscure a fuller understanding of the important features arising from thermal effects.

In this Letter we report a $T$ dependent study of the magnetization reversal in individual submicron Py disks of different sizes performed via micro-Hall \cite{Rahm2003, Li2002} and bend-resistance \cite{Hara2006} magnetometry. Although for intermediate-size disks we observe complex reversal behavior, preventing distinct conclusion about the role of thermal energy in nucleation and annihilation of magnetic vortices, we were able to identify thermal effects in the smallest and largest disks. We find that at low temperatures vortex nucleation and annihilation proceed via thermal activation over an energy barrier, while at high temperatures they are governed by the $T$ dependence of the saturation magnetization of the disk.

\begin{figure}
\includegraphics[scale=0.54, bb=0 0 450 397]{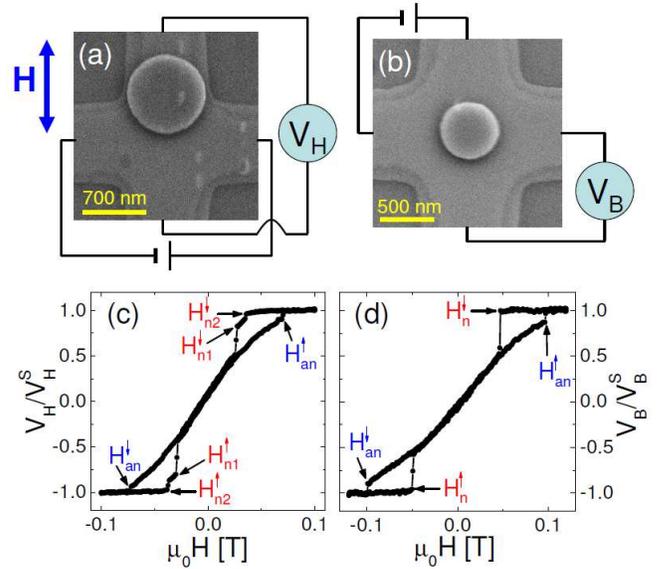}
\caption{(Color online) Scanning electron microscope images of Hall cross magnetometers and Py disks adapted to show the measurement configuration for: (a) Hall magnetometry for 865-nm disk; (b) bend resistance magnetometry for 526-nm disk. (c),(d) Hysteresis loops measured at 20~K for the disks shown in (a) and (b), respectively.}
\label{fig:SEM+hloops}
\end{figure}

Hall cross magnetometers were fabricated from a molecular beam epitaxy grown InAs quantum well semiconductor heterostructure consisting of a GaAs substrate/GaAs buffer (100~nm)/AlAs (10~nm)/AlSb
(30~nm)/Al$_{0.7}$Ga$_{0.3}$Sb (1000~nm)/AlSb (8~nm) /InAs (12.5~nm)/AlSb (13~nm)/GaSb (0.6~nm)/In$_{0.5}$Al$_{0.5}$As (5~nm). Micro-Hall devices fabricated from this material have been previously utilized in detection \cite{Mihajlovic2005, Manandhar2009} and magnetic characterization \cite{MihajlovicAPL2007} of superparamagnetic beads, and their characteristics are described  in Ref.~\onlinecite{MihajlovicJAP2007}. Here, Hall crosses with an active area of 1~$\times$~1~$\mu$m$^2$ and contact leads were defined by e-beam and optical lithography, respectively, followed by wet chemical etching. In a subsequent e-beam lithography step four disks (diameters 526, 643, 744  and 865~nm) were patterned using a double layer PMMA resist, e-beam evaporation of 50-nm thick Py, and subsequent lift-off. Only one Py disk was patterned per cross [see Figs. 1(a) and (b)] to enable single-disk measurements. The hysteresis loops for each disk were measured by applying an in-plane magnetic field $H$ and recording either the resulting Hall voltage $V_H$ [see Fig.~1(a)] or the bend voltage $V_B$ [see Fig.~1(b)] while biasing the devices with a $dc$ current of 20 $\mu$A.

Figure 1(d) shows a hysteresis loop measured at 20~K for the 526-nm disk shown in Fig.~1(b). The measured $V_B$ data are normalized to the saturation value $V_B^S$. The loop shows well defined $H_n$ and $H_{an}$ values with a linear region in between, corresponding to reversible vortex displacement. In contrast, the  reversal of the 865-nm disk shown in Fig.~1(a) is more complex. Figure 1(c) shows the hysteresis loop measured for this disk at 20~K. Here, sharp changes in the magnetization due to the vortex nucleation, $H_{n1}$, are preceded by small kinks at larger fields, $H_{n2}$. This behavior has been observed previously in micro-Hall \cite{Rahm2003} and magnetic force microscopy \cite{Pokhil2000} measurements on individual Py disks, and was attributed to the development of an $S$-shape magnetization configuration and a double-vortex, prior to single vortex formation \cite{Guslienko2001}. The double-to-single vortex magnetization reversal is more favorable in larger disks \cite{Pokhil2000, Guslienko2001}, consistent with our observation.

\begin{figure}
\includegraphics[scale=0.43, bb=18 15 560 236]{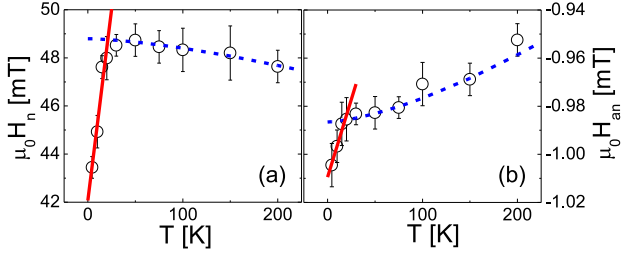}
\caption{(Color online) $T$ dependencies of (a) $H_n$ and (b) $H_{an}$, for the 526-nm disk. The solid red and dashed blue line in both figures correspond to fittings based on models of thermal activation over an energy barrier and saturation magnetization driven vortex formation, respectively.}
\label{fig:resistance}
\end{figure}

Figures 2(a) and (b) show the $T$ dependence of $H_n$ and $H_{an}$ for the 526-nm disk. The data are obtained as $H_n = (H_n^\downarrow - H_n^\uparrow)/2$ and $H_{an} = (H_{an}^\downarrow - H_{an}^\uparrow)/2$, where "$\downarrow$" and "$\uparrow$" correspond to decreasing and increasing field $H$, respectively. Variations of both critical fields with $T$ are clearly observed. $H_n$ exhibits a non-monotonic $T$ dependence with a sharp increase on warming below 50~K and a slow decrease above it. On the other hand, the annihilation field monotonically increases with decreasing temperature, but with distinctly different slopes in the low and the high $T$ regions. Also, the position of the maximum in the $T$ dependence of $H_n$ coincides with the $T$ value at which the change in slope of $H_{an}~vs.~T$ occurs.

\begin{figure}
\includegraphics[scale=0.43, bb=18 15 553 413]{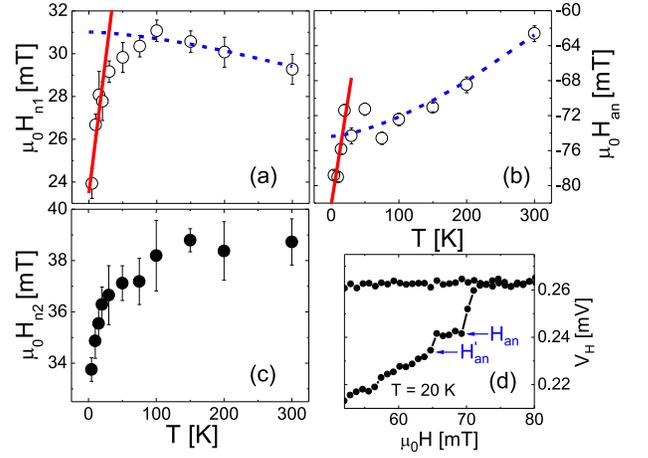}
\caption{(Color online) $T$ dependencies of (a) $H_{n1}$, (b) $H_{an}$, and (c) $H_{n2}$ for the 865-nm disk. The solid red and dashed blue lines in Figs. (a) and (b) correspond to fittings based on models of thermal activation over an energy barrier, and saturation magnetization driven vortex formation, respectively. (d) Portion of the hysteresis loop for the 865-nm disk showing double-step like annihilation of a magnetic vortex.}
\label{fig:resistance}
\end{figure}

While an increase of $H_n$ with increasing $T$ below $\sim$50~K, is consistent with thermally activated vortex nucleation over an energy barrier, the decreasing of $H_n$  above
$\sim$50~K is surprising. It suggests a non-physical scenario in which the energy barrier for vortex nucleation increases with increasing $T$. However, this observation can be explained by the decrease of the saturation magnetization $M_s$ of the disk due to thermally excited spin waves, since $H_n$ is proportional to $M_s$. This is also consistent with the overall $T$ dependence of  $H_{an}$: at low $T$ the vortex persists longer with decreased thermal activation towards the uniform state, while at high $T$ the decrease of $M_s$ facilitates vortex annihilation. The dashed blue lines in Figs.~2(a) and (b) show the fit to $H_n$ and $H_{an}$ data points respectively, above 50~K, to the formula $H_{n,an} = H_{(n,an)0}(1-\alpha_{n,an} T^{3/2})$, expected based on the thermal dependence of $M_s$. We find
$\alpha_n = (0.8 \pm 0.2) \times 10^{-5}$~K$^{-3/2}$ and
$\alpha_{an} = (1.0 \pm 0.2) \times 10^{-5}$~K$^{-3/2}$, which agree well with each other and also with the literature $\alpha$ value for thin Py films. At low $T$, based on thermal activation over an energy barrier, and assuming a linear dependence of the energy barrier height on the external magnetic field, we expect $H_{n,an} = H_{(n,an)0}(1 + \beta_{n,an} T)$. Fitting this formula to the data below 20~K [see red solid lines in Figs. 2(a) and (b)] we find
$\beta_n = (7.5 \pm 1.4) \times 10^{-3}$~K$^{-1}$ and
$\beta_{an} = (1.3 \pm 0.2) \times 10^{-3}$~K$^{-1}$, consistent with results for Py squares \cite{Li2001}, considering the different shapes. However, it remains unclear why there is a distinct transition between the two mechanisms at low and high $T$. We point out that the overall $T$ dependence of $H_n$ and $H_{an}$ cannot be quantitatively described considering both mechanisms over the whole temperature-range. Further studies should be performed to elucidate this issue.

The general behavior of $H_n$ and $H_{an}$ presented for the 526-nm disk was also observed for the 865-nm disk. Figures 3(a) and (c) show $T$ dependencies of $H_{n1}$ and $H_{n2}$ [see Fig.~1(c)]. For $H_{n1}$ we find a similar increase at low $T$ and a slow decrease at high $T$ with
$\alpha_n = (1.0 \pm 0.3) \times 10^{-5}$~K$^{-3/2}$ and
$\beta_n = (10.8 \pm 3.8) \times 10^{-3}$~K$^{-1}$ obtained from fittings (dashed blue and solid red curve, respectively), in good agreement with the the 526-nm disk. On the other hand, $H_{n2}$ does not decrease at high $T$. This suggests that the demagnetizing fields do not have a strong effect on the critical field for the formation of a double-vortex. From $H_{an}$ shown in Fig.~3(b) we find
$\alpha_{an} = (3.0 \pm 0.5) \times 10^{-5}$~K$^{-3/2}$ and
$\beta_{an} = (5.9 \pm 1.9) \times 10^{-3}$~K$^{-1}$. These values disagree slightly with the previous ones, possibly because of the less consistent nature of the vortex annihilation observed in this disk. Namely, in some cases the annihilation of a magnetic vortex occurred via a double-step transition, as illustrated in Fig.~3(d) for 20~K data,  suggesting that even vortex annihilation may proceed via different reversal modes.

\begin{figure}
\includegraphics[scale=0.63, bb=0 0 365 385]{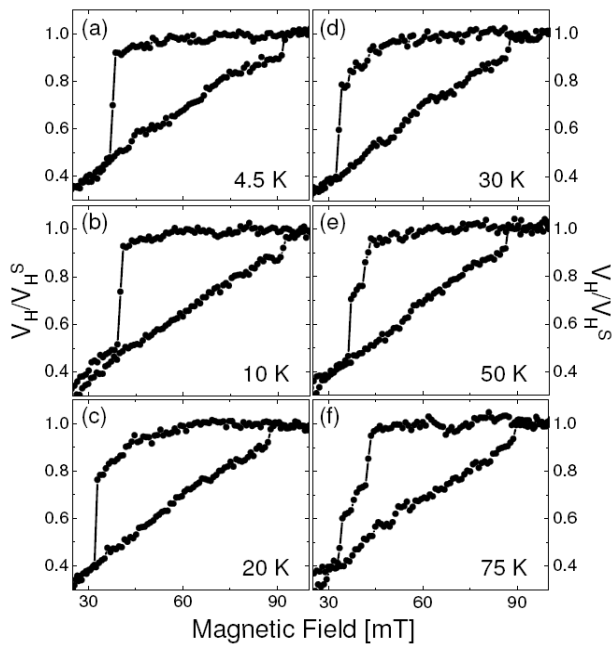}
\caption{(a)-(f) Portion of the hysteresis loops for the 644-nm disk at different $T$ values.  The data show that magnetization reversal occurs via different modes at the different $T$ values.}
\label{fig:resistance}
\end{figure}

While the qualitative behavior did not vary with $T$ for these two disks, the situation was more complex for the intermediate-sized disks. This can be seen in Figs.~4(a)-(f), where details of the vortex nucleation for the 644-nm disk are shown at six different $T$ values between 4.5 and 75~K. For the two lowest temperatures [see Figs.~4(a) and (b)] only one sharp transition from the uniform state is observed, suggesting single vortex formation from an almost perfectly uniform state. When $T$ is increased, the $V_H$ is continuously reduced before nucleation [see Figs.~4(c) and (d)], which suggests the formation of a $C$-like state prior to vortex nucleation. Further increasing $T$ leads to reversal via formation of a double-vortex [see Figs.~4(e) and (f)], similar to the case of the largest disk. The variety of possible reversal modes that may appear even in a single disk as a function of temperature suggests that any valid $T$ dependent studies of magnetization reversal can be performed only by single disk measurements.

In conclusion, we studied magnetization reversal in individual submicrometer Py disks with different diameters. We observed a clear $T$ dependence of the critical fields for vortex nucleation and annihilation in the smallest and largest disk, which suggest that two different mechanisms, namely vortex formation and annihilation over an energy barrier and decrease in saturation magnetization dominate the $T$ dependencies at low and high $T$, respectively. While these two mechanisms considered separately can explain the data even in a quantitative way, the overall $T$ dependence cannot be explained by considering these two together. In intermediate-sized disks we found that the magnetization reversal can be complex, with $T$ dependent transitions between single-vortex, $C$ state, and double-vortex states.

The work at Argonne was supported by the U.S.\ DOE-BES, under contract No. DE-AC02-06CH11357.


\end{document}